\documentclass[times,onecolumn]{aastex62}

\usepackage{cases}
\usepackage{graphicx}
\usepackage{subfigure}
\usepackage{natbib}
\usepackage{color}
\usepackage{tabularx}
\usepackage{bm}
\usepackage{threeparttable}

\newcommand{\thetae}{\theta_{\rm E}}
\newcommand{\pie}{\pi_{\rm E}}

\newcommand{\te}{t_{\rm E}}
\newcommand{\event}{TCP J0507+2447}

\newcommand{\Sp}{{\it Spitzer}}

\newcommand{\mplanet}{M_{\rm planet} = 19.0 \pm 3.0~M_{\earth}}

\DeclareGraphicsExtensions{.pdf,.png,.jpg}

\shorttitle{}
\shortauthors{Zang et al.}

\begin{document}

\title{{\large \Sp\ + VLTI-GRAVITY Measure the Lens Mass of a Nearby Microlensing Event}}

\correspondingauthor{Weicheng Zang, Subo Dong}
\email{zangwc17@mails.tsinghua.edu.cn, dongsubo@pku.edu.cn}
\author[0000-0001-6000-3463]{Weicheng Zang}
\affiliation{Department of Astronomy and Tsinghua Centre for Astrophysics, Tsinghua University, Beijing 100084, China}
\author{Subo Dong}
\affiliation{Kavli Institute for Astronomy and Astrophysics, Peking University, Yi He Yuan Road 5, Hai Dian District, Beijing 100871, China}
\author{Andrew Gould}
\affiliation{Max-Planck-Institute for Astronomy, K\"onigstuhl 17, 69117 Heidelberg, Germany}
\affiliation{Department of Astronomy, Ohio State University, 140 W. 18th Ave., Columbus, OH 43210, USA}
\author{Sebastiano Calchi Novati}
\affiliation{IPAC, Mail Code 100-22, Caltech, 1200 E. California Blvd., Pasadena, CA 91125, USA}
\author{Ping Chen}
\affiliation{Kavli Institute for Astronomy and Astrophysics, Peking University, Yi He Yuan Road 5, Hai Dian District, Beijing 100871, China}
\affiliation{Department of Astronomy, School of Physics, Peking University, Yi He Yuan Road 5, Hai Dian District, Beijing 100871, China}
\author{Hongjing Yang}
\affiliation{Department of Astronomy and Tsinghua Centre for Astrophysics, Tsinghua University, Beijing 100084, China}
\author[0000-0001-9952-7408]{Shun-Sheng Li}
\affiliation{National Astronomical Observatories, Chinese Academy of Sciences, Beijing 100101, China}
\affiliation{School of Astronomy and Space Science, University of Chinese Academy of Sciences, Beijing 100049, China}
\author{Shude Mao}
\affiliation{Department of Astronomy and Tsinghua Centre for Astrophysics, Tsinghua University, Beijing 100084, China}
\affiliation{National Astronomical Observatories, Chinese Academy of Sciences, Beijing 100101, China}
\author{K.B. Alton}\affiliation{Desert Blooms Observatory}
\author{J. Brimacombe}\affiliation{Coral Towers Observatory, Cairns, Queensland 4870, Australia}
\author{Sean Carey}
\affiliation{Spitzer Science Center, MS 220-6, California Institute of Technology,Pasadena, CA, USA}
\author{G.~W. Christie}\affiliation{Auckland Observatory, Box 24180, Auckland, New Zealand}
\author{F.~Delplancke-Str\"obele}\affiliation{European Southern Observatory, Karl-Schwarzschild-Str. 2, 85748 Garching, Germany}
\author{Dax~L. Feliz}\affiliation{Department of Physics and Astronomy, Vanderbilt University, Nashville, TN 37235, USA}
\author{J. Green}\affiliation{Kumeu Observatory, Kumeu, New Zealand}
\author{Shaoming Hu}\affiliation{Shandong Provincial Key Laboratory of Optical Astronomy and Solar-Terrestrial Environment, Institute of Space Sciences, Shandong University, Weihai 264209, People's Republic of China}
\author{T. Jayasinghe}\affiliation{Department of Astronomy Ohio State University, 140 W.\ 18th Ave., Columbus, OH 43210, USA} \affiliation{Center for Cosmology and AstroParticle Physics (CCAPP), The Ohio State University, 191 W. Woodruff Avenue, Columbus, OH 43210, USA.}
\author{R.~A. Koff}\affiliation{Antelope Hills Observatory, 980 Antelope Drive West, Bennett, CO 80102, USA}
\author{A. Kurtenkov}\affiliation{Institute of Astronomy and National Astronomical Observatory, Bulgarian Academy of Sciences, 72
Tsarigradsko Shose Blvd., 1784 Sofia, Bulgaria}
\author{A.~M\'erand}\affiliation{European Southern Observatory, Karl-Schwarzschild-Str. 2, 85748 Garching, Germany}
\author{Milen Minev}\affiliation{Department of Astronomy, Faculty of Physics, University of Sofia, 1164 Sofia, Bulgaria}
\author{Robert Mutel}\affiliation{Department of Physics and Astronomy, University of Iowa}
\author{T. Natusch}\affiliation{Institute for Radio Astronomy and Space Research, AUT University, Auckland, New Zealand}
\author{Tyler Roth}\affiliation{Department of Physics and Astronomy, University of Iowa}
\author{Yossi Shvartzvald}\affiliation{Department of Particle Physics and Astrophysics, Weizmann Institute of Science, Rehovot 76100, Israel}
\author{Fengwu Sun}\affiliation{Steward Observatory, University of Arizona, 933 North Cherry Avenue, Tucson, AZ 85721, USA}
\author{T. Vanmunster}\affiliation{Center for Backyard Astrophysics Belgium, Walhostraat 1A, B-3401 Landen, Belgium}
\author{Wei Zhu}\affiliation{Canadian Institute for Theoretical Astrophysics, University of Toronto, 60 St George Street, Toronto, ON M5S 3H8, Canada}


\begin{abstract}
We report the lens mass and distance measurements of the nearby microlensing event TCP J05074264+2447555. We measure the microlens parallax vector $\bm{\pi}_{\rm E}$ using \Sp\ and ground-based light curves with constraints on the direction of lens-source relative proper motion derived from Very Large Telescope Interferometer (VLTI) GRAVITY observations. Combining this $\bm{\pi}_{\rm E}$ determination with the angular Einstein radius $\thetae$ measured by VLTI GRAVITY observations, we find that the lens is a star with mass $M = 0.495 \pm 0.063~M_{\odot}$ at a distance $D_{\rm L} = 429 \pm 21~{\rm pc}$. We find that the blended light basically all comes from the lens. The lens-source proper motion is $\mu_{\rm rel,hel} = 26.55 \pm 0.36~{\rm mas\,yr^{-1}}$, so with currently available adaptive-optics (AO) instruments, the lens and source can be resolved in 2021. This is the first microlensing event whose lens mass is unambiguously measured by interferometry + satellite parallax observations, which opens a new window for mass measurements of isolated objects such as stellar-mass black holes.

\end{abstract}

\section{Introduction}\label{intro}

Gravitational microlensing offers a unique window into probing extrasolar planets beyond the snow line \citep{Shude1991,Andy1992,Mao2012,Gaudi2012} and isolated dark objects such as free-floating planets \citep{Sumi2011, Mroz2017a, MrozNeptune,Mroz2FFP}, brown dwarfs \citep{OB07224,OB170896} and black holes \citep{Gouldremnant,ShudeBH,DaveBH,OGLE3BH}. For a typical microlensing event, the only measured observable that relates to the physical properties of the lens is the Einstein timescale $\te$. It is a combination of the lens mass $M_{\rm L}$, the lens-source relative proper motion $\bm{\mu}_{\rm rel}$ and parallax $\pi_{\rm rel}$ by 
\begin{equation}\label{mass}
    \te = \frac{\thetae}{\mu_{\rm rel}}; \qquad \thetae = \sqrt{\kappa M_{\rm L} \pi_{\rm rel}}; \qquad \pi_{\rm rel} = \pi_{\rm L} - \pi_{\rm S}, 
\end{equation}
where $\kappa \equiv 4G/(c^2\mathrm{AU}) = 8.144$ mas$/M_{\odot}$, $\thetae$ is the angular Einstein radius, $\pi_{\rm L} = {\rm AU}/D_{\rm L}$ and $\pi_{\rm S} = {\rm AU}/D_{\rm S}$ are the lens and source parallax, respectively, and $D_{\rm L}$ and $D_{\rm S}$ are the lens and the source distances, respectively. Therefore, with only $\te$ known, the lens mass and distance cannot be unambiguously determined. By far, the most common way to break this degeneracy is to also measure the angular Einstein radius $\thetae$ and the microlens parallax $\bm{\pi}_{\rm E}$. For a lensing object, its mass is related to these two observables by \citep{Gould1992, Gould2000}
\begin{equation}
    M_{\rm L} = \frac{\thetae}{{\kappa}\pie},
    \label{eq:mass}
\end{equation}
and the lens-source relative proper motion and parallax by 
\begin{equation}
 \bm{\mu}_{\rm rel} = \frac{\thetae}{\te}\frac{\bm{\pi}_{\rm E}}{\pie}; \qquad \pi_{\rm rel} = \thetae\pie,
\end{equation}
where the microlens parallax vector $\bm{\pi}_{\rm E}$ is defined by
\begin{equation}\label{piemu}
    \bm{\pi}_{\rm E} \equiv \frac{\pi_{\rm rel}}{\thetae} \frac{\bm{\mu}_{\rm rel}}{\mu_{\rm rel}}.
\end{equation}


There are three methods to measure the microlens parallax $\pie$. The first one is ``orbital microlens parallax'', which is due to the Earth's orbital acceleration around the Sun that introduces deviation from rectilinear motion in the lens-source relative motion \citep{Gould1992}. However, this method is generally only feasible for events with long microlensing timescales $\te\gtrsim$ year/$2\pi$ \citep[e.g.,][]{OB171434} and/or events produced by nearby lenses \citep[e.g.,][]{OB170537}. 
The second method is ``terrestrial microlens parallax'' \citep{HardyWalker1995, HolzWald1996}, which in rare cases can be measured by a combination of simultaneous observations from well-separated ground-based telescopes \citep[e.g.,][]{OB07224,OB080279}. The most efficient and robust way to measure $\bm{\pi}_{\rm E}$ is via ``satellite microlens parallax'', which is done by observing the same microlensing event from Earth and one or more well-separated ($\sim$ AU) satellite \citep{1966MNRAS.134..315R,1994ApJ...421L..75G,Gould1995single}. The first observed example applying this method was the event OGLE-2005-SMC-001 \citep{OB05001}, in which the joint analysis of ground-based observations and \Sp\ observations indicates that the lens is probably a halo binary. Microlens parallax measurements from two satellites (\Sp\ and the two-wheel {\it Kepler} K2) have also been achieved to measure the parallax of the event OGLE-2016-BLG-0975 \citep{Zhu2017K2Spitzer}. Since 2014, the \Sp\ satellite has observed about 1000 microlensing events toward the Galactic bulge \citep{GouldSp1,GouldSp2,GouldSp3,GouldSp4,GouldSp5,GouldSp6}, in order to probe planets in substantially different Galactic environments \citep{Novati2015,Zhu2017spitzer}, and they have yielded unambiguous mass measurements for seven planetary systems \citep{OB140124,OB150966,OB161190,OB161195,OB171140,OB180596,KB180029}. 

The angular Einstein radius $\thetae$ is generally measured via finite-source effects \citep{1994ApJ...421L..75G,Shude1994,Nemiroff1994} when the source transits a caustic (where the magnification diverges to infinity) or comes close to a cusp. The detection of finite-source effects usually yields the source radius normalized by the Einstein radius, $\rho$, and combining it with the source angular radius $\theta_*$, which is routinely determined from the intrinsic color and de-reddened brightness of the source \citep{Yoo2004}, can lead to the measurement of the angular Einstein radius $\thetae = \theta_*/\rho$. Finite-source effects are frequently detected in binary/planetary events due to their relatively large caustic structures, but they are rarely measured in a single-lens event since its caustic is a single geometric point.

Besides the mass measurements from combining the angular Einstein radius $\thetae$ and the microlens parallax $\pie$, an independent mass-distance relationship can be obtained if the flux from the lens system is measured with high angular resolution imaging and compared to stellar models \citep[see, e.g.,][]{DaveAO,JenniferAO}. These are achieved either by measuring the flux at the position of the source star in excess of the source flux to constrain the lens flux \citep[e.g.,][]{MB16227,OB120950_1}, or in some cases, resolving the source and lens $\sim$ 5--20 years after the microlensing event and thus directly measure the lens flux \citep[e.g.,][]{Alock2001, Batista2015,Dave2015,OB120950}. The additional mass-distance relationship from the lens flux, combined with the constraints from the angular Einstein radius $\thetae$ and/or the microlens parallax, can yield the mass and distance of the lens system \citep[e.g.,][]{OB050071D,Beaulieu2018}. However, this method is not feasible for dark lenses such as free-floating planets and black holes.     

Here we report the lens mass measurement of the nearby microlensing event TCP J05074264+2447555 (hereafter referred to as ``\event'' for brevity) by a joint analysis of ground-based, \Sp, and VLTI GRAVITY observations of \cite{Jan2017}. The paper is structured as follows. In \S~\ref{obser}, we introduce the ground-based and \Sp\ observations. We then describe the light curve modeling process in \S~\ref{model}. In \S~\ref{lens}, we derive the physical parameters of the lens. Finally, we discuss the implications of our work in \S~\ref{dis}.


\section{Observations}\label{obser}

\event, at equatorial coordinates $(\alpha, \delta)_{\rm J2000}$ = (05:07:42.72, $+24$:47:56.4, \citealt{Gaia2018AA}), corresponding to Galactic coordinates $(\ell,b)=(178.76, -9.33)$, was first discovered by the Japanese amateur astronomer Tadashi Kojima (Gunma-ken, Japan) on UT 2017-10-25.688. For ground-based data, we choose the light curves used by \cite{Jan2017}, including the data from All-Sky Automatic Survey for Supernovae (ASAS-SN; \citealt{ASASSN}), 0.6 m telescopes at Post Observatory (RP), 0.5 m Iowa Robotic Telescope (Iowa) at the Winer Observatory (Arizona, USA), 0.4m telescope at Auckland Observatory (AO), and the 1.3m SMARTS telescope \citep{CT13} at Cerro Tololo Inter-American Observatory (CTIO). We supplement it with CTIO $V$-band data to derive the $VHL$ color-color relation. All the ground-based data were calibrated to standard magnitude systems. For further descriptions of our ground-based data and their availability in digital format, see Appendix \S~\ref{sec:ground}.

For \Sp, we submitted a Director Discretion Time (DDT) proposal \citep{spitzerddt} to observe the \event\,on 2017 November 7, and it was approved on 2017 November 9. Due to the Sun-angle limit, it did not start taking observations until 2017 December 19 (${\rm HJD}^{\prime} = 8107.2, {\rm HJD}^{\prime} = {\rm HJD} - 2450000$). The observations ended on 2018 January 23 (${\rm HJD}^{\prime} = 8143.7$). In total, 21 data points were taken, all using the 3.6 $\mu$m channel ($L-$band) of the IRAC camera. The data were reduced by the method presented by \cite{Spitzerdata}.

\section{Light curve analysis}\label{model}

\begin{figure}[htb] 
    \includegraphics[width=\columnwidth]{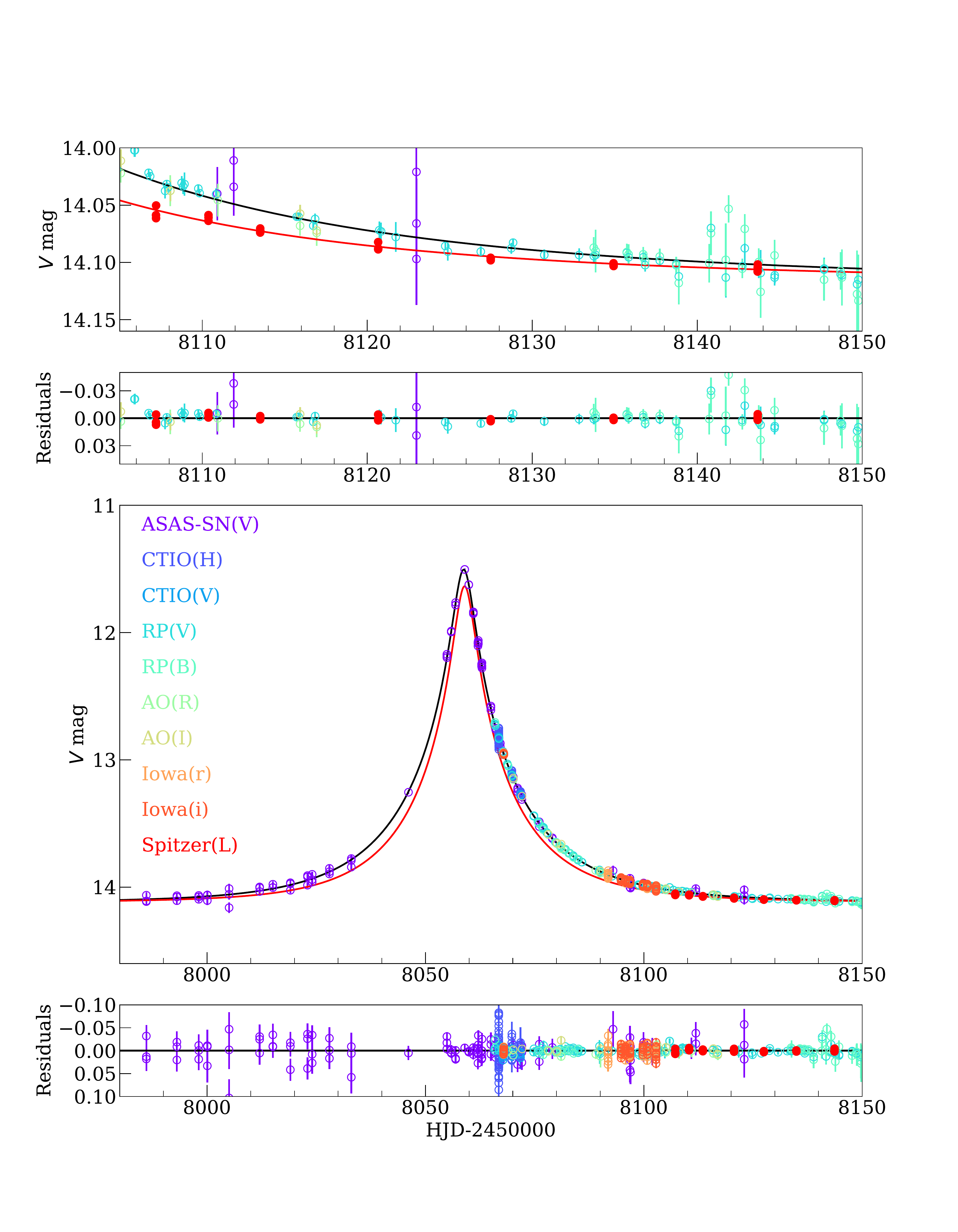}
    \caption{The light curves and data points of \event. The upper panel show a close-up of \Sp\ data points. The black line represent the best-fit $u_{0,\earth} > 0$ with lens light model for the ground data, and the red line shows the corresponding model for \Sp. The circles with different colors are ground-based data points from different telescopes or bands. The red dots are \Sp\ data points.}
    \label{lc}
\end{figure}

Figure \ref{lc} shows the \event\ data together with the best-fit single-lens model. In this section, we analyze the data with a single-lens model. \cite{Jan2017_2} reported a short-duration planetary anomaly near the peak of the event, and we discuss the planetary-lens modeling in Appendix\S~\ref{sec:planet_model}, in order to double check whether the measurements of the microlens parallax $\pie$ is affected by the planetary model.

\subsection{Ground-based data only}\label{ground}
\begin{figure}[htb] 
    \includegraphics[width=12.5cm]{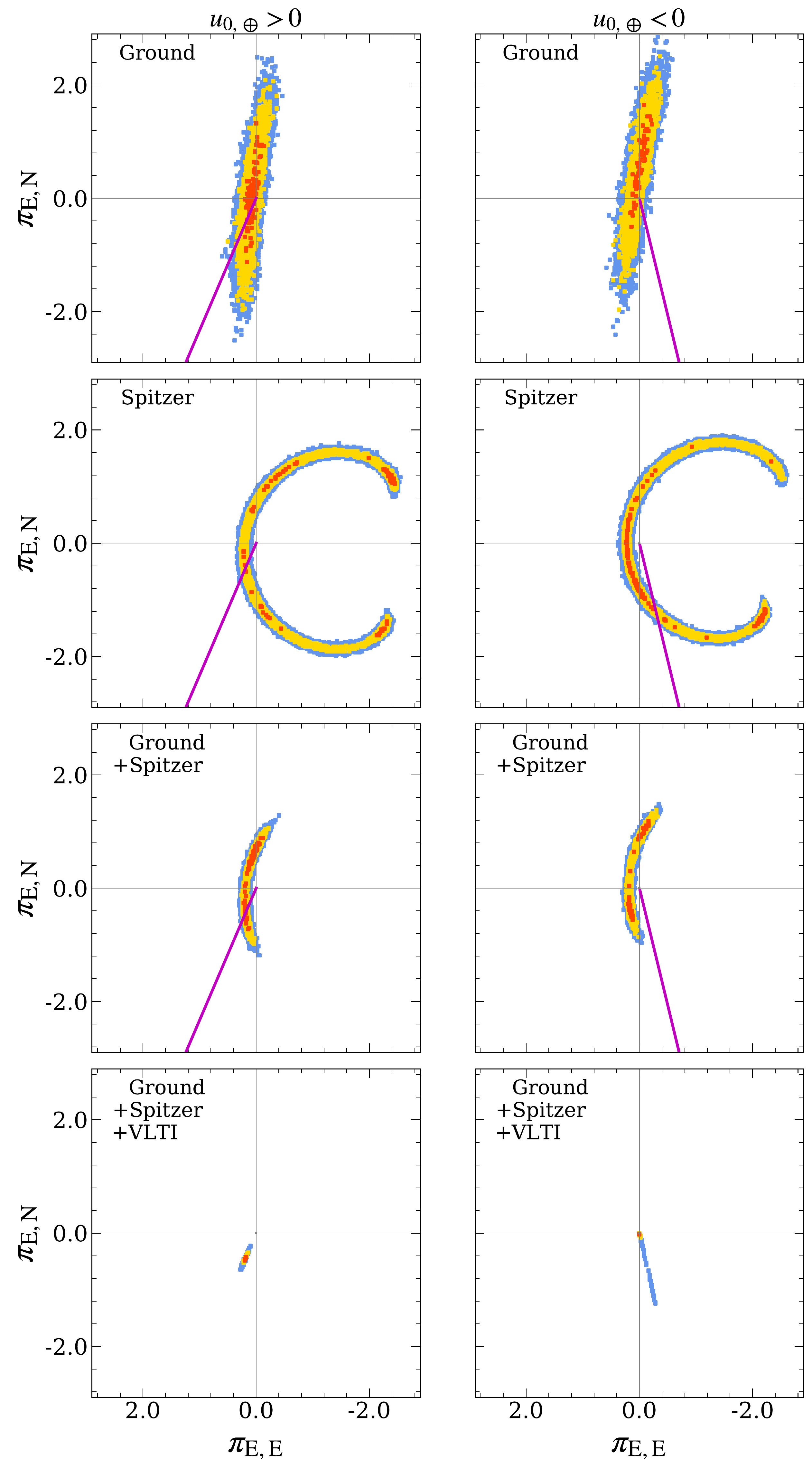}
    \centering
    \caption{Likelihood distributions for $\bm{\pi_{\rm E}}$ derived from MCMC. The left and right panels show the distributions for $u_0 > 0$ and $u_0 < 0$ solutions, respectively. Red, yellow, and blue show likelihood ratios $[-2\Delta\ln{\mathcal{L}/\mathcal{L}_{\rm max}}] < (1, 4, \infty)$, respectively. The magenta lines represent the best value of the VLTI directions.}
    \label{pie}
\end{figure}

The single-lens model has three parameters $t_0$, $u_0$, $\te$ \citep{Paczynski1986} to calculate the magnification as a function of time $A(t)$: the time of the maximum magnification $t_0$, the impact parameter $u_0$ (in units of the angular Einstein radius $\thetae$), and the Einstein radius crossing time $\te$. For each data set $i$, we introduce two flux parameters ($f_{{\rm S},i}$, $f_{{\rm B},i}$) in order to model the observed flux $f_{i}(t)$ as
\begin{equation}
    f_{i}(t) = f_{{\rm S},i} A(t) + f_{{\rm B},i},
\end{equation}
where $f_{{\rm S},i}$ represents the flux of the source star, and $f_{{\rm B},i}$ represents any blended flux that is not lensed in the photometric aperture. In addition, to fit the orbital microlens parallax, we parameterize the microlens parallax effects by $\pi_{\rm E,N}$ and $\pi_{\rm E,E}$, which are the North and East components of the microlens parallax vector \citep{Gouldpies2004}, respectively. We also fit $u_{0,\earth} > 0$ and $u_{0,\earth} < 0$ solutions to consider the ``ecliptic degeneracy'' \citep{Jiang2004, Poindexter2005}. We find $\pie = 0.13\pm0.47$ for the $u_{0,\earth} > 0$ solution and $\pie = 0.67\pm0.52$ for the $u_{0,\earth} < 0$ solution. The likelihood distributions of $(\pi_{\rm E,N}, \pi_{\rm E,E})$ from ground-based data are shown in Figure \ref{pie}.

\subsection{Satellite parallax}\label{parallax}
For \event, \Sp\ took observations in $1.7 < (t-t_{0,\earth})/\te < 3.1$. We can estimate the microlens parallax by
\begin{equation}\label{equ:para}
\vec{\pi}_{\rm E} \sim \frac{\rm au}{D_{\perp}}\left(\Delta\tau,\Delta\beta\right),\qquad \Delta\tau\equiv\frac{t_{0,Spitzer}-t_{0,\rm\oplus}}{t_{\rm E}},\qquad \Delta\beta\equiv\pm u_{0,Spitzer}-\pm u_{0,\earth},
\end{equation}
where $D_\perp$ is the projected separation between the \Sp\ satellite\footnote{We extract the geocentric locations of \Sp\ from the {\it JPL Horizons} website: \url{http://ssd.jpl.nasa.gov/?horizons}} and Earth at the time of the event. Generally, the four possible values of $\Delta\beta$ result in a set of four degenerate solutions (\citealt{1966MNRAS.134..315R}; see also Figure 1 from \citealt{1994ApJ...421L..75G}). However, the four solutions merge into two disjoint ($u_{0,\earth} > 0$ and $u_{0,\earth} < 0$) solutions because the \Sp\ data commence well after $t_{0,Spitzer}$ \citep{Gould2019}. In addition, we include a $VHL$ color-color constraint on the \Sp\ source flux $f_{s,Spitzer}$ \citep[e.g.,][]{Novati2015}, which adds a $\chi^2_{\rm Color}$ into the total $\chi^2_{\rm total}$,
\begin{equation}
    \chi^2_{\rm Color} = \frac{[(V - L)_{\rm S} - (V - L)_{\rm fix}]^2}{\sigma_{\rm cc}^2},
\end{equation}
where $(V - L)_{\rm S}$ is the source color from the modeling, $(V - L)_{\rm fix}$ is the color constraint, and $\sigma_{\rm cc}$ is the uncertainty of the color constraint. To derive the color constraint, we extract the CTIO $V$- and $H$- band and \Sp\ $L$-band photometry for stars within $1'$ (10 stars in total), and fit for the two parameters $c_0$ and $c_1$ in the equation
\begin{equation}
    V_{\rm CTIO} - L_{\rm \Sp} = c_0 + c_1(V_{\rm CTIO} - H_{\rm CTIO} - X_p),
\end{equation}
where $X_p = 3.14$ is a pivot parameter chosen to minimize the covariance between $c_0$ and $c_1$. We get $c_0 = 2.370 \pm 0.016, c_1 = 0.988 \pm 0.043$. We derive the source color by regression of CTIO $V$ versus $H$ flux as the source magnification changes, and find $(V_{\rm CTIO} - H_{\rm CTIO})_{\rm S} = 2.096 \pm 0.024$, $(V - L)_{\rm fix} = 1.34 \pm 0.06$. We apply the color constraint to the modeling and find $\pie = 0.68\pm0.23$ for the $u_{0,\earth} > 0$ solution and $\pie = 1.01\pm0.34$ for the $u_{0,\earth} < 0$ solution. 

The measured satellite parallax above is a combination of orbital parallax and satellite parallax. To investigate the satellite parallax from the \Sp\ data, we fix $(t_0, u_{0,\earth}, \te)$ from the best-fit non-parallax model, and fit the parallax only with \Sp\ data and the color constraint. We find the resulting ``Ground + {\it Spitzer}'' parallax is an intersection of the satellite parallax, which is nearly a circle, and the orbital parallax, which is nearly a straight line. \cite{Gould2019} shows that the nearly circular shape of the satellite-parallax contours are due to partial overlap of a series of osculating, exactly circular, degeneracies in the $\bm{\pi}_{\rm E}$ plane, and similar satellite-parallax shapes have been detected in two \Sp\ planetary events \citep{OB180596,KB180029}. The best-fit parameters for the $u_{0,\earth} > 0$ and $u_{0,\earth} < 0$ solutions are shown in Tables \ref{parm1} and \ref{parm2}, respectively. The likelihood distributions of $(\pi_{\rm E,N}, \pi_{\rm E,E})$ from ``{\it Spitzer}-only'' and ``Ground + {\it Spitzer}'' are shown in Figure \ref{pie}.   

\subsection{VLTI constraint on the parallax direction}
\begin{figure}[htb] 
    \includegraphics[width=12.5cm]{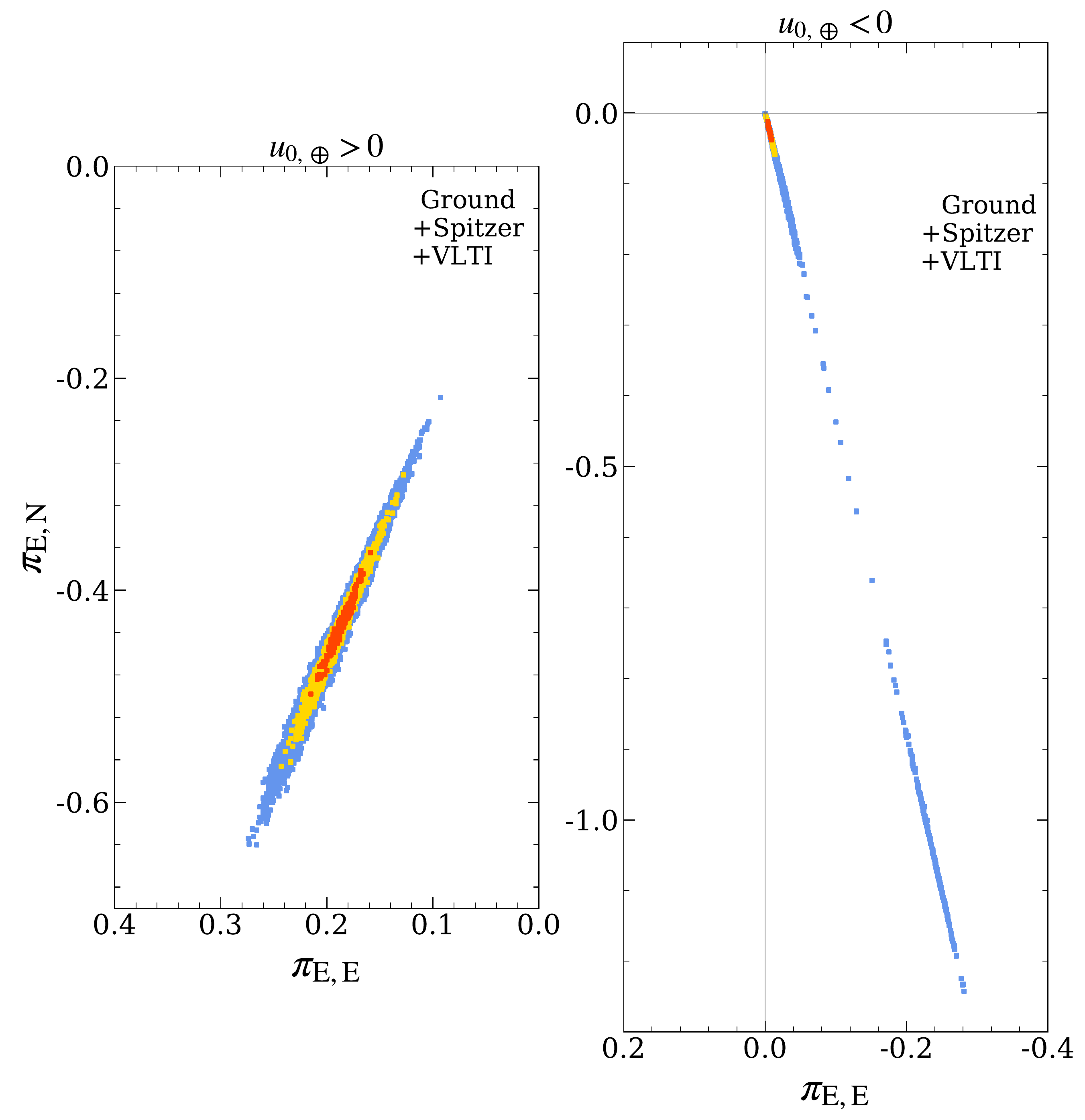}
    \centering
    \caption{The Close-up of $\bm{\pi_{\rm E}}$ Likelihood distributions with Ground + \Sp\ + VLTI constraints. The symbols are similar to those in Figure \ref{pie}.}
    \label{pie_small}
\end{figure}

VLTI GRAVITY provides constraints on the direction of the lens-source relative proper motion and thus (see Equation \ref{piemu}) the direction of the microlens parallax (see Figure 4 of \citealt{Jan2017}). For the ``no lens light'' model, the direction of the microlens parallax  (North through East)
\begin{numcases}{\Phi_{\rm VLTI} =}
192.9^{\circ} \pm 0.4^{\circ}~{\rm for}~u_{0,\earth} < 0 \\
156.1^{\circ} \pm 0.4^{\circ}~{\rm for}~u_{0,\earth} > 0,      
\end{numcases}
and for the ``luminous lens '' model,
\begin{numcases}{\Phi_{\rm VLTI} =}
193.5^{\circ} \pm 0.4^{\circ}~{\rm for}~u_{0,\earth} < 0 \\
156.7^{\circ} \pm 0.4^{\circ}~{\rm for}~u_{0,\earth} > 0.      
\end{numcases}
We include the constraint of the parallax direction by adding a $\chi^2_{\rm VLTI}$ into the total $\chi^2_{\rm total}$,
\begin{equation}
    \chi^2_{\rm VLTI} = \frac{(\Phi_{\rm model} - \Phi_{\rm VLTI})^2}{\sigma_{\rm VLTI}^2},
\end{equation}
where $\Phi_{\rm VLTI}$ is the parallax direction from the modeling, and $\sigma_{\rm VLTI} = 0.4^{\circ}$ is the uncertainty of the VLTI parallax direction. The best-fit parameters are shown in Tables \ref{parm1} and \ref{parm2}, and its likelihood distributions of $(\pi_{\rm E,N}, \pi_{\rm E,E})$ are shown in Figure \ref{pie_small}. For both the ``no lens light'' model and the ``luminous lens'' model, the $u_{0,\earth} < 0$ solution is disfavored by $\Delta\chi^2 > 41$. Actually, from the likelihood distributions of $(\pi_{\rm E,N}, \pi_{\rm E,E})$ from ``Ground + {\it Spitzer}'' data shown in Figure \ref{pie}, we can see that the VLTI parallax direction for the $u_{0,\earth} < 0$ solution is inconsistent with the constraint from the ``Ground + {\it Spitzer}'' data. Thus, we reject the $u_{0,\earth} < 0$ solution.

\section{Physical parameters}\label{lens}

\subsection{Is the lens luminous?}\label{luminous}
For the ``no lens light'' model, VLTI measured the angular Einstein radius $\thetae = 1.850 \pm 0.014$ mas, and the light curve modeling shows $\pie = 0.476 \pm 0.061$, so the lens mass $M_{\rm L} = 0.477 \pm 0.061 M_{\odot}$ and the lens source relative parallax $\pi_{\rm rel} = \thetae\pie = 0.880 \pm 0.113~{\rm mas}$. The parallax of the ``baseline object'' has been measured by {\it Gaia} second data release (DR2) \citep{Gaia2018AA, Gaia_dis}
\begin{equation}\label{pibase}
  \pi_{\rm base} = 1.480 \pm 0.031~{\rm mas}  
\end{equation}
which is the flux-weighted mean parallax of the source and the blend in {\it Gaia} band,
\begin{equation}\label{piweight}
  \pi_{\rm base} = \eta\pi_{\rm B} + (1 - \eta)\pi_{\rm S},
\end{equation}
where $\eta$ is the fraction of total {\it Gaia} flux due to the blending, $\pi_{\rm B}$ is the parallax of the blending. The {\it Gaia} band is qualitatively similar to $V$ band, and the best-fit solution of the ``no lens light'' model has $f_{{\rm B},V}/f_{{\rm S},V} \sim 0.04 $, so we estimate $\pi_{\rm S} \sim \pi_{\rm base}$, and $D_{\rm L} = {\rm AU}/(\pi_{\rm rel} + \pi_{\rm S}) = 424 \pm 14~{\rm pc}$. Using the MIST isochrones\footnote{\url{http://waps.cfa.harvard.edu/MIST/interp isos.html}} with age $\geq$ 2Gyr, the lens has an absolute $K$-band magnitude of $K_{\rm L,0} = 5.7 \pm 0.3$. \cite{Jan2017} has found that the source suffers from an extinction $A_{K,{\rm S}} = 0.155$, we thus estimate the extinction of the lens $A_{K,{\rm L}} \sim A_{K,{\rm S}}D_{\rm L}/D_{\rm S} = 0.097$. As a result, the apparent $K$-band magnitude of the lens is 
\begin{equation}
   K_{\rm L} = K_{\rm L,0} + 5 \log \frac{D_{\rm L}}{10~{\rm pc}} + A_{K,{\rm L}} = 13.9 \pm 0.3,
\end{equation}
which is $\sim 13\%$ of the 2MASS baseline $K = 11.680 \pm 0.018$. Thus, the lens is luminous, and we reject the ``no lens light'' model.

\subsection{Blend = Lens?}

Table \ref{blend} shows the blend in $H$-, $I$-, $R$-, $V$- bands from the best-fit ``luminous lens'' model, which shows that the blend is detected in all bands. To estimate the lens apparent brightness, we use the angular Einstein radius $\thetae = 1.891 \pm 0.014$ mas and $\pie = 0.469 \pm 0.060$ of the ``luminous lens'' model, and follow the procedure in Section \ref{luminous}. For extinction, we adopt the extinction law of \cite{Cardelli89} and estimate the extinction of the lens in $\lambda$ band $A_{\lambda,{\rm L}} \sim A_{\lambda,{\rm S}}D_{\rm L}/D_{\rm S}$. The predicted lens apparent magnitude is shown in Table \ref{blend}. We find that the predicted lens apparent magnitude is consistent with the blend within $1\sigma$ in all bands. Thus, the blended light basically all comes from the lens.

\subsection{Lens parameters}
According to the best-fit ``luminous lens'' model, the lens mass 
\begin{equation}
    M = \frac{\thetae}{\kappa\pie} = 0.495 \pm 0.063~M_{\odot},
\end{equation}
the lens-source relative parallax
\begin{equation}
    \pi_{\rm rel} = \thetae\pie = 0.887 \pm 0.114~{\rm mas}.
\end{equation}
Combining Equations (\ref{mass}), (\ref{pibase}) and (\ref{piweight}), we obtain
\begin{equation}
    \pi_{\rm L} = \pi_{\rm base} + (1 - \eta)\pi_{\rm rel}  = 2.332 \pm 0.114~{\rm mas}; \qquad D_{\rm L} = \frac{\rm AU}{\pi_{\rm L}} = 429 \pm 21~{\rm pc},
\end{equation}
where we adopt $\eta = f_{{\rm B},V}/(f_{{\rm S},V}+f_{{\rm B},V}) = 0.040 \pm 0.016$ from the best-fit ``luminous lens'' model as the fraction of total {\it Gaia} flux due to the lens. The geocentric and heliocentric relative proper motion are
\begin{equation}
    \bm{\mu_{\rm rel,geo}}(N, E) = \frac{\thetae}{\te}\frac{\bm{\pi}_{\rm E}}{\pie} = (-22.73, 9.83) \pm (0.21, 0.19)~{\rm mas\,yr^{-1}}; 
\end{equation}
\begin{equation}
    \bm{\mu_{\rm rel,hel}}(N, E) = \bm{\mu_{\rm rel,geo}} + \frac{\pi_{\rm rel}}{{\rm AU}}\bm{v}_{\earth,\perp} = (-22.45, 14.18) \pm (0.21, 0.59)~{\rm mas\,yr^{-1}},
\end{equation}
where $\bm{v}_{\earth,\perp}=(1.47,23.29)$~km~s$^{-1}$ is Earth's projected velocity on the event at $t_0$. The proper motion of the ``baseline object'' has been measured by {\it Gaia}
\begin{equation}
    \bm{\mu}_{\rm base}(N, E) = (-7.330, -0.228) \pm (0.033, 0.061)~{\rm mas\,yr^{-1}},
\end{equation}
which is also the flux-weighted mean proper motion. Thus, the lens and source proper motion are
\begin{equation}
    \bm{\mu}_{\rm L}(N, E) = \bm{\mu}_{\rm base} + (1 - \eta)\bm{\mu_{\rm rel,hel}} = (-28.89, 13.39) \pm (0.41, 0.61)~{\rm mas\,yr^{-1}};
\end{equation}
\begin{equation}
    \bm{\mu}_{\rm S}(N, E) = \bm{\mu}_{\rm base} - \eta\bm{\mu_{\rm rel,hel}} = (-6.43, -0.80) \pm (0.36, 0.23)~{\rm mas\,yr^{-1}}.
\end{equation}
We summarize the derived lens parameters in Table \ref{lensparm}. 


\section{Discussion and Conclusion}\label{dis}

We have reported the analysis of the microlensing event \event. The combination of the angular Einstein radius $\thetae$ measured by VLTI GRAVITY observations, and the direction of microlens parallax $\bm{\pi}_{\rm E}$ constrained from VLTI GRAVITY observations, \Sp\ observations and ground-based observations, reveals that the lens is a $M_{\rm L} = 0.495 \pm 0.063~M_{\odot}$ star at $D_{\rm L} = 429 \pm 21~{\rm pc}$. We also note that \cite{Kojima2} reported the lens mass and distance of \event\ to be $M_{\rm L} = 0.581 \pm 0.033~M_{\odot}$ star at $D_{\rm L} = 505 \pm 47~{\rm pc}$, which is a combination of finite-source effects, Keck AO image, spectroscopy, annual parallax and VLTI observations (see Table 4 and Figure 9 of \citealt{Kojima2}). That is, their mass estimate is $1.2\,\sigma$ higher than ours. We note that \cite{Kojima2} used spectroscopy to determine the source distance and got $D_{\rm S} = 800 \pm 130~{\rm pc}$. If we adopt {\it Gaia} parallax, the angular Einstein radius $\thetae$ measured by VLTI GRAVITY observations and the lens flux constraints of \cite{Kojima2}, we get $M_{\rm L} = 0.527 \pm 0.032~M_{\odot}$ star at $D_{\rm L} = 434 \pm 12~{\rm pc}$, which is in agreement with our measurements within $1\sigma$. Using current AO instruments, \cite{Batista2015,Dave2015,OB120950} resolved the lens and source for cases that these have approximately equal brightness when they were separated by $34$--$60$ mas. In this case, the lens-source proper motion is $\mu_{\rm rel,hel} = 26.55 \pm 0.36~{\rm mas\,yr^{-1}}$, and the lens is about $1.8$ mag fainter than the source in $H$-band. We estimate that it will probably require $\sim 80$ mas to resolve the source and lens. Thus, the derived physical parameters of our work can be verified by currently available AO instruments in 2021 and later. 

This is the first microlensing event whose lens mass has been unambiguously measured by interferometry (VLTI GRAVITY) + satellite ({\it Spitzer}) parallax observations. Interferometry + satellite parallax is a new method to measure the mass of isolated objects. The detection frequency of finite-source effects in a single event is only $\sim 2\%$ \citep{ZhuPLFS,PLFS}, so interferometric observation such as VLTI GRAVITY is a complementary approach to measurements of the angular Einstein radius $\thetae$. In addition, interferometry can provide additional constraints on the microlens parallax direction, which is helpful for breaking degeneracy in the parallax measuremnets \citep{1966MNRAS.134..315R,1994ApJ...421L..75G}. \cite{Gouldremnant} estimated that $\sim20\%$ of Galactic microlensing events are caused by stellar remnants, and specifically that $\sim1\%$ are due to stellar-mass black-hole lenses. Some black-hole candidates \citep{ShudeBH,DaveBH,OGLE3BH} have been reported by microlensing surveys, but none of these candidates had finite-source effects. Therefore, interferometry observations such as VLTI GRAVITY + satellite parallax opens up a new window for decisively confirming black-hole candidates by measuring their masses.

\cite{Jan2017_2} reported a planetary companion in the \event\ lens system. Using the derived lens parameters, we infer the planet mass to be
\begin{equation}
    \mplanet.
\end{equation}
The projected planet-host separation is
\begin{numcases}{a_{\perp} = s\thetae D_{\rm L}}
0.76 \pm 0.04 ~{\rm AU}~{\rm for}~ s \sim 0.935; \\
0.79 \pm 0.04 ~{\rm AU}~{\rm for}~ s \sim 0.975,      
\end{numcases}
where $s$ is the planet-host projected separation in units of $\thetae$. Following the procedure of \cite{OB180740}, we estimate the radial-velocity (RV) amplitude $v\sin(i) \sim 2.3~{\rm m}~{\rm s}^{-1}$ with a period of $\sim 1.3$ yr, which may be detectable by high-resolution spectrometers such as VLT/Espresso with 4 VLT telescopes. In addition, the snow line radius of the lens system is $a_{\rm SL} \sim 2.7(M/M_{\odot}) = 1.3$~{\rm AU} \citep{snowline}. Thus, this is the first Neptune within the snow line detected by microlensing.


\acknowledgments
We thank Tianshu Wang and Jennifer Yee for fruitful discussions. We are grateful to Robin Leadbeater and Paolo Berardi for making their spectroscopic observations available to us during the observing campaign. S.D. and P.C. acknowledges Projects 11573003 supported by the National Science Foundation of China (NSFC). W.Z., H.Y., S.-S.L. and S.M. acknowledges support by the National Science Foundation of China (Grant No. 11821303 and 11761131004). Work by AG was supported by AST-1516842 and by JPL grant 1500811. AG received support from the European Research Council under the European Unions Seventh Framework Programme (FP 7) ERC Grant Agreement n. [321035]. This work is based (in part) on observations made with the Spitzer Space Telescope, which is operated by the Jet Propulsion Laboratory, California Institute of Technology under a contract with NASA. Support for this work was provided by NASA through an award issued by JPL/Caltech. Work by S. M. Hu was supported by the Natural Science Foundation of Shandong province (No. JQ201702), and the Young Scholars Program of Shandong University (No. 20820162003). WZ was supported by the Beatrice and Vincent Tremaine Fellowship at CITA.

\appendix
\section{the Planetary Model}
\label{sec:planet_model}
\cite{Jan2017_2} reported a short-duration anomaly near the peak of the event, indicating that the lens star has a planetary companion with planet-host mass ratio $q = 1.1 \times 10^{-4}$. To double check the parallax measurements, we fit the event with the binary-lens model. Besides the three parameters $t_0$, $u_0$, $\te$ introduced in \S~\ref{ground}, the binary-lens model has four other parameters ($s, q, \alpha, \rho$). Here, $q$ is the companion-host mass ratio, $s$ is the companion-host projected separation in units of $\thetae$, $\alpha$ is the angle between the source trajectory and the binary axis in the lens plane, and $\rho$ is the source radius normalized to $\thetae$. 

We fix ($s, q, \alpha$) as the best-fit values shown in \cite{Jan2017_2}. We include the satellite parallax and the constraints from VLTI GRAVITY. We consider the so-called ``minor-image perturbation degeneracy'' \citep[e.g.,][]{OB120950_1,OB161067,MB16319} found by \cite{Jan2017_2}: $s = 0.935 \pm 0.004$ and $s = 0.975 \pm 0.004$. Table \ref{parm3} shows the best-fit parameters. For all the solutions, the resulting parallax is consistent with the results of the single-lens model within 1$\sigma$. Thus, different models (single-lens and binary-lens) do not have significant influence on the parallax and thus the mass measurements.

\section{Ground-based Photometric Observations}
\label{sec:ground}

We summarize the ground-based photometric observations collected during our observing campaign in Table~\ref{tab:sites}. The ASCII data will be made available upon publication.

\begin{table*}[htbp]
\label{tab:sites}
\caption{Ground-based Photometric Observations}
\begin{center}
\begin{tabular}{cccccccc}
\hline
\hline
 Site    &  Filter \\
 \hline
 All-Sky Automatic Survey for Supernovae (ASAS-SN)     & $V$   \\ 
 Cerro Tololo Inter-American Observatory (CTIO) & $V$, $H$ \\
 Post Observatory (RP) & $B$, $V$ \\
 Auckland Observatory (AO) & $R$, $I$ \\
 Iowa Robotic Telescope (Iowa) & $r$, $i$ \\
 Desert Bloom Observatory (DBO) & $B$, $V$, $I$, Clear \\
 Coral Towers Observatory (CTO) & Clear \\
 Kumeu Observatory (KO) & Wratten12 \\
 Weihai (WH) Observatory of Shandong University & $B$, $V$ \\
 Antelope Hill Observatory (ATO) &  $B$, $V$ \\
 Bulgarian National Astronomical Observatory (Rozhen) & $B$, $V$\\
 Peking University 40cm Telescope of  (PFT) & UHC \\
 Center for Backyard Astronomy Belgium Observatory (CBABO) & $V$, Clear\\
\hline 
\end{tabular}
\end{center}
\end{table*}


\bibliography{Zang.bib}

\begin{table}[htb]
    \centering
    \caption{Best-fit parameters and their $68\%$ uncertainty range from MCMC for the $u_{0,\earth} > 0$ solutions}
    \begin{tabular}{c|c c c c c}
    \hline
    \hline
    Parameters  & \multicolumn{5}{c}{$u_{0,\earth} > 0$} \\
    \hline
       & non-parallax  & w/o \Sp\ & w/ \Sp\ & VLTI + \Sp\ (w/o lens light) & VLTI + \Sp\ (w/ lens light) \\
    \hline
    $t_{0,\earth}$ (${\rm HJD}^{\prime}$) & 8058.76(1) & 8058.76(1) & 8058.76(1) & 8058.76(1) & 8058.76(1) \\
    $u_{0,\earth}$  & 0.083(1) & 0.084(1) & 0.084(1) & 0.085(1) & 0.084(1) \\
    $\te$ & 28.25(15) & 27.98(42) & 27.89(32) & 27.83(15) & 27.89(15)  \\
    $\pi_{\rm E,N}$ & - & 0.11(83) & 0.68(56) & $-0.435(56)$ & $-0.430(55)$ \\
    $\pi_{\rm E,E}$ & - & 0.06(13) & 0.01(11) & 0.193(24) & 0.186(24)\\
    $\pi_{\rm E}$ & - & 0.13(47) & 0.68(23) & 0.476(61) & 0.469(60) \\
    $V_{\rm CTIO,S}$ & 14.18(1) & 14.17(2) & 14.17(2) & 14.15(1)  & 14.16(1)\\
    $V_{\rm CTIO,B}$ & 17.17(10) & 17.38(32) & 17.36(27)  & 17.58(16) & 17.55(15)\\
    $H_{\rm CTIO,S}$ & 12.04(1) & 12.03(2) & 12.03(2) & 12.02(1) & 12.02(1) \\
    $H_{\rm CTIO,B}$ & 13.69(3) & 13.74(8) & 13.74(7) & 13.79(3) & 13.78(3) \\
    $\chi^2_{\rm Color}$ & - & - & 0.040 & 0.002 & 0.047 \\
    $\chi^2_{\rm VLTI}$ & - & - & - & 0.014 & 0.003 \\
    $\chi^2_{\rm total}/dof$ & 557.82/558 & 556.9/556 & 576.2/575 & 576.7/575 & 576.6/575 \\
    \hline
    \hline
    \end{tabular}
    \label{parm1}
\end{table}

\begin{table}[htb]
    \centering
    \caption{Best-fit parameters and their $68\%$ uncertainty range from MCMC for the $u_{0,\earth} < 0$ solutions}
    \begin{tabular}{c|c c c c c}
    \hline
    \hline
    Parameters  & \multicolumn{5}{c}{$u_{0,\earth} < 0$} \\
    \hline
      & non-parallax  &  w/o \Sp\ & w/ \Sp\ & VLTI + \Sp\ (w/o lens light) & VLTI + \Sp\ (w/ lens light) \\
    \hline
    $t_{0,\earth}$ (${\rm HJD}^{\prime}$) & 8058.76(1) & 8058.76(1) & 8058.76(1) & 8058.76(1) & 8058.76(1) \\
    $u_{0,\earth}$ & $-$0.083 & $-$0.084(1)  & $-$0.084(1) & $-$0.084(1) & $-$0.082(1) \\
    $\te$ & 28.26(16) & 28.11(45) & 28.24(38) & 28.06(35) & 28.15(29) \\
    $\pi_{\rm E,N}$ & - & 0.67(81) & 1.01(63) & $-$0.003(234) & $-$0.003(223) \\
    $\pi_{\rm E,E}$ & - & $-$0.01(16) &  $-0.07$(13) & $-$0.001(50) & $-$0.001(51) \\
    $\pi_{\rm E}$ & - & 0.67(52) & 1.01(34) & 0.003(239) & 0.003(229) \\
    $V_{\rm CTIO,S}$ & 14.18(1) & 14.16(2) & 14.17(2) & 14.17(1) & 14.17(1) \\
    $V_{\rm CTIO,B}$ & 17.17(10) & 17.40(34) & 17.33(29) & 17.27(27) & 17.22(23) \\
    $H_{\rm CTIO,S}$ & 12.04(1) & 12.02(2) & 12.03(2) & 12.03(1) & 12.04(1)  \\
    $H_{\rm CTIO,B}$ & 13.69(3) & 13.75(8) & 13.73(7) & 13.71(6) & 13.70(6) \\
    $\chi^2_{\rm Color}$ & - & - & 0.002 & 39.840 & 40.728 \\
    $\chi^2_{\rm VLTI}$ & - & - & -  & 0.033 & 0.009 \\
    $\chi^2_{\rm total}/dof$ & 557.83/558 & 556.7/556 & 576.2/575 & 618.2/575 & 618.0/575 \\
    \hline
    \hline
    \end{tabular}
    \label{parm2}
\end{table}

\begin{table}[htb]
    \centering
    \caption{Blend from the best-fit ``luminous lens'' model versus the predicted apparent magnitude of the lens}
    \begin{tabular}{c c c c c}
    \hline
    \hline
    Band & $H$  & $I$ & $R$ & $V$ \\
    \hline
    Extinction $A_{\lambda}$ & 0.10 & 0.42 & 0.66 & 0.87 \\
    Blending & $13.78 \pm 0.03$ & $16.30 \pm 0.07$ & $16.76 \pm 0.10$ & $17.55 \pm 0.15$ \\
    Predicted lens apparent magnitude  & $13.9 \pm 0.3$  & $16.2 \pm 0.4$ & $16.7 \pm 0.6$ & $17.6 \pm 0.8$ \\
    
    \hline
    \hline
    \end{tabular}\\
    \label{blend}
\end{table}

\begin{table}[htb]
    \centering
    \caption{Physical parameters for the lens of \event}
    \begin{tabular}{c|c}
    \hline
    $M_{\rm L}$ [$M_{\odot}$] & $0.495 \pm 0.063$ \\
    $D_{\rm L}$ [pc] & $429 \pm 21$ \\
    $D_{\rm S}$ [pc] & $692 \pm 15$ \\
    $\mu_{\rm L,hel,N}$ [$\rm mas\,yr^{-1}$] & $−28.89 \pm 0.41$ \\
    $\mu_{\rm L,hel,S}$ [$\rm mas\,yr^{-1}$] & $13.39 \pm 0.61$  \\
    $\mu_{\rm S,hel,N}$ [$\rm mas\,yr^{-1}$] & $-6.42 \pm 0.36$  \\
    $\mu_{\rm S,hel,S}$ [$\rm mas\,yr^{-1}$] & $-0.80 \pm 0.23$  \\ 
    $\mu_{\rm rel,hel, N}$ [$\rm mas\,yr^{-1}$] & $-22.45 \pm 0.21$ \\
    $\mu_{\rm rel,hel, E}$ [$\rm mas\,yr^{-1}$] & $14.18 \pm 0.59$ \\
    $\mu_{\rm rel,geo, N}$ [$\rm mas\,yr^{-1}$] & $-22.73 \pm 0.21$ \\
    $\mu_{\rm rel,geo, E}$ [$\rm mas\,yr^{-1}$] & $9.83 \pm 0.19$ \\

    \hline
    \hline
    \end{tabular}\\
    \label{lensparm}
\end{table}

\begin{table}[htb]
    \centering
    \caption{Binary-lens fitting results  with the VLTI constraints}
    \begin{tabular}{c|c c c c| c c c c}
    \hline 
    Parameters & \multicolumn{4}{c|}{$s = 0.935$} & \multicolumn{4}{c}{$s = 0.975$} \\
    \hline
        & \multicolumn{2}{c|}{$u_{0,\earth} > 0$} & \multicolumn{2}{c|}{$u_{0,\earth} < 0$} & \multicolumn{2}{c|}{$u_{0,\earth} > 0$} & \multicolumn{2}{c}{$u_{0,\earth} < 0$} \\
    \hline
        & w/o lens light & w/ lens light & w/o lens light &  w/ lens light & w/o lens light & w/ lens light & w/o lens light &  w/ lens light \\
    \hline
    $t_{0,\earth}$ (${\rm HJD}^{\prime}$) & 8058.76(1) & 8058.76(1) & 8058.76(1) & 8058.76(1) & 8058.76(1) & 8058.76(1) & 8058.78(1) & 8058.76(1) \\
    $u_{0,\earth}$  & 0.086(1) & 0.086(1) & $-0.094(1)$ & $-0.086(3)$ & 0.086(1) & 0.086(1) & $-0.079$ & $-0.084$ \\
    $\te$ & 27.82(16) & 27.82(16) & 28.10(41) & 28.04(67) & 27.81(17) & 27.84(16) & 29.61(48) & 28.15(37) \\
    $q (10^{-4})$  & 1.1 & 1.1 & 1.1 & 1.1 & 1.1 & 1.1 & 1.1 & 1.1 \\
    $\alpha$ (rad) & 1.625 & 1.625 & 4.658 & 4.658 & 1.516 & 1.516 & 4.767 & 4.767 \\
    $\rho(10^{-3})$ & 5.5(6) & 5.5(6) & 81(16) & 5.6(8) & 5.8(6) & 6.1(6) & 5.1(5) & 6.1(6) \\
    $\pi_{\rm E,N}$ & $-0.427(57)$ & $-0.433(56)$ & $-0.014(337)$ & $-0.023(435)$ & $-0.445(56)$ & $-0.432(58)$ & $-0.004(292)$ & $-0.004(287)$ \\
    $\pi_{\rm E,E}$ & 0.189(25) & 0.188(23) & $-0.003(76)$ & $-0.006(102)$ & 0.197(24) & 0.186(25) & $-0.001(66)$ & $-0.001(65)$ \\
    $\pi_{\rm E}$ & 0.467(62) & 0.472(61) & 0.015(346) & 0.024(447) & 0.486(61) & 0.470(63) & 0.004(299) & 0.004(295) \\
    $V_{\rm CTIO,S}$ & 14.15(1) & 14.15(1) & 14.17(2) & 14.16(3) & 14.15(1) & 14.15(1) & 14.17(3) & 14.17(3)  \\
    $V_{\rm CTIO,B}$ & 17.71(20) & 17.68(19) & 17.38(41) & 17.39(62) & 17.70(20) & 17.71(20) & 17.29(52) & 17.28(53) \\
    $H_{\rm CTIO,S}$ & 12.02(1) & 12.02(1) & 12.03(2) & 12.03(3) & 12.02(1) & 12.02(1) & 12.03(3) & 12.03(3) \\
    $H_{\rm CTIO,B}$ & 13.81(4) & 13.81(4) & 13.74(10) & 13.74(14) & 13.81(4) & 13.81(4) & 13.72(13) & 13.71(13) \\
    $\chi^2_{\rm Color}$ & 0.026 & 0.019 & 41.722 & 41.891 & 0.027 & 0.058 & 41.643 & 41.274 \\
    $\chi^2_{\rm VLTI}$ & 0.040 & 0.114 & 0.074 & 0.028 & 0.024 & 0.026 & 0.202 & 0.390 \\
    $\chi^2_{\rm total}/dof$ & 575.1/554 & 575.2/554 & 618.9/574 & 619.0/574 & 575.3/574 & 575.2/574 & 617.8/574 & 617.2/574 \\
    \hline
    \hline
    \end{tabular}
    \label{parm3}
\end{table}

\end{document}